\documentclass[12pt]{article}
\usepackage{amssymb}
\newcommand{\be}{\begin{eqnarray}}
\newcommand{\ee}{\end{eqnarray}}
\newcommand{\non}{\nonumber}
\newcommand{\tr}{\mathop{\rm tr}\nolimits}
\newcommand{\id}{\mathbb{I}}
\newcommand{\C}{\mathbb{C}}

\begin{document}

\begin{titlepage}
\strut\hfill UMTG--214
\vspace{.5in}
\begin{center}

\LARGE A Spin Chain Primer \\[1.0in]
\large Rafael I. Nepomechie\\[0.8in]
\large Physics Department, P.O. Box 248046, University of Miami\\[0.2in]  
\large Coral Gables, FL 33124 USA\\

\end{center}

\vspace{.5in}

\begin{abstract}
This is a very elementary introduction to the Heisenberg (XXX) quantum 
spin chain, the Yang-Baxter equation, and the algebraic Bethe Ansatz.
\end{abstract}

\end{titlepage}

\section{Introduction}

Like the quantum harmonic oscillator, the Heisenberg quantum spin 
chain is one of the fundamental models of physics.  Indeed, it is 
simply formulated, it describes real systems which are experimentally 
accessible \footnote{See, e.g., F.  Essler's contribution to these 
Proceedings.}, it has a rich and elegant mathematical structure, it is 
the prototype of all integrable models, and it has close connection to 
integrable and conformal field theory.  Here we review the formulation 
of the model, and we describe the elements of the so-called algebraic 
Bethe Ansatz approach for diagonalizing the Hamiltonian.  The aim of 
these notes is to enable beginning graduate students to start 
exploring the vast literature on this subject.

\section{Heisenberg spin chain}

We formulate the spin $1/2$ isotropic Heisenberg (XXX) quantum spin 
chain first for one site, then for two sites, and then finally for $N$ 
sites.

\subsection{One site}

Actually, the one-site problem is too trivial: one cannot write down 
the Heisenberg spin Hamiltonian for only one site.  Nevertheless, it 
is important to identify the set of observables and the vector space 
on which these operators act.

For the one-site problem, a basis for the observables consists of the 
identity matrix $\id$ and the familiar Pauli spin matrices
$\vec \sigma$, that is
\be
\id = \left( \begin{array}{cc}
	               1  & 0 \\
				   0  & 1 
\end{array} \right) \,, \qquad  
\sigma^{x}=\left( \begin{array}{cc}
	               0  & 1 \\
				   1  & 0 
\end{array} \right) \,, \qquad  
\sigma^{y}=\left( \begin{array}{cc}
	               0  & -i \\
				   i  & 0 
\end{array} \right) \,, \qquad 
\sigma^{z}=\left( \begin{array}{cc}
	               1  & 0 \\
				   0  & -1 
\end{array} \right) \,.
\ee
These operators act on a 
two-dimensional complex vector space $V = \C^{2}$, whose 
elements we represent by two-component vectors 
$x = \left( \begin{array}{c}
      x_{1} \\
      x_{2}
\end{array} \right)$, with $x_{i} \in \C$.

\subsection{Two sites}

We generalize to the case of more than one site using
the notion of tensor product. For the $2 \times 2$ matrices 
$A= \left( \begin{array}{cc}
	               a_{11}  & a_{12} \\
				   a_{21}  & a_{22}
\end{array} \right)$ and 
$B= \left( \begin{array}{cc}
	               b_{11}  & b_{12} \\
				   b_{21}  & b_{22}
\end{array} \right)$, the tensor product $A \otimes B$ is defined 
by \footnote{I like to draw the horizontal and vertical lines to help 
organize the matrix elements; they have no other significance, and 
can be omitted.}

\be
\left( \begin{array}{cc}
	               a_{11}  & a_{12} \\
				   a_{21}  & a_{22}
\end{array} \right) \otimes
\left( \begin{array}{cc}
	               b_{11}  & b_{12} \\
				   b_{21}  & b_{22}
\end{array} \right) = \left( \begin{array}{cc|cc}
a_{11} b_{11} & a_{11} b_{12} & a_{12} b_{11} & a_{12} b_{12}  \\ 
a_{11} b_{21} & a_{11} b_{22} & a_{12} b_{21} & a_{12} b_{22}  \\
\hline
a_{21} b_{11} & a_{21} b_{12} & a_{22} b_{11} & a_{22} b_{12}  \\
a_{21} b_{21} & a_{21} b_{22} & a_{22} b_{21} & a_{22} b_{22}  
\end{array} \right) \,.
\ee 

For the two-site problem, the basic observables are the spin operators 
at each site, $\vec \sigma_{1} \equiv \vec \sigma \otimes \id$ 
and $\vec \sigma_{2} \equiv \id \otimes \vec \sigma$.  
That is, \footnote{For visual clarity, I write only the nonzero matrix elements
in the tensor product.}
\be
\sigma_{1}^{x} &=& \left( \begin{array}{cc}
	               0  & 1 \\
				   1  & 0 
\end{array} \right) \otimes \left( \begin{array}{cc}
	               1  & 0 \\
				   0  & 1 
\end{array} \right) = \left( \begin{array}{cc|cc}
 &  &  1  & \\ 
 &  &  &  1  \\
\hline
 1  &  &  & \\
 &  1  &  &
\end{array} \right) \,, \non \\ 
\sigma_{1}^{y} &=& \left( \begin{array}{cc}
	               0  & -i \\
				   i  & 0 
\end{array} \right) \otimes \left( \begin{array}{cc}
	               1  & 0 \\
				   0  & 1 
\end{array} \right) = \left( \begin{array}{cc|cc}
 &  &  -i &  \\ 
 &  &  & -i \\
\hline
 i  &  &  & \\
 &  i  &  &
\end{array} \right) \,, \non \\ 
\sigma_{1}^{z} &=& \left( \begin{array}{cc}
	               1  & 0 \\
				   0  & -1 
\end{array} \right) \otimes \left( \begin{array}{cc}
	               1  & 0 \\
				   0  & 1 
\end{array} \right) = \left( \begin{array}{cc|cc}
 1 &  &  &  \\ 
 & 1  &  &  \\
\hline
 & & -1  & \\
 & &  & -1 
\end{array} \right) \,,
\ee 
and 
\be 
\sigma_{2}^{x} &=& \left( \begin{array}{cc}
	               1  & 0 \\
				   0  & 1 
\end{array} \right) \otimes
\left( \begin{array}{cc}
	               0  & 1 \\
				   1  & 0 
\end{array} \right)   = \left( \begin{array}{cc|cc}
 &  1  &  & \\ 
 1  &  &  & \\
\hline
 &  &  & 1  \\
 &  &  1 &
\end{array} \right) \,, \non \\ 
\sigma_{2}^{y} &=& \left( \begin{array}{cc}
	               1  & 0 \\
				   0  & 1 
\end{array} \right)  \otimes
\left( \begin{array}{cc}
	               0  & -i \\
				   i  & 0 
\end{array} \right) = \left( \begin{array}{cc|cc}
 & -i  &  & \\ 
 i  &  &  & \\
\hline
 & & & -i   \\
 & & i  &
\end{array} \right) \,, \non \\ 
\sigma_{2}^{z} &=& \left( \begin{array}{cc}
	               1  & 0 \\
				   0  & 1 
\end{array} \right) \otimes \left( \begin{array}{cc}
	               1  & 0 \\
				   0  & -1 
\end{array} \right) = \left( \begin{array}{cc|cc}
 1  &  &  & \\ 
 & -1  &  & \\
\hline
 &  &  1 &  \\
 &  &  & -1 
\end{array} \right) \,.
\ee 
Note that 
\be
\left[ \sigma_{1}^{i} \,, \sigma_{2}^{j} \right] = 0
\ee
for any $i \,, j \in \{ x \,, y \,, z \}$.

These operators act on the tensor product space $V \otimes V$, with 
elements 
\be
x \otimes y = \left( \begin{array}{c}
      x_{1} \\
      x_{2}
\end{array} \right) \otimes
\left( \begin{array}{c}
      y_{1} \\
      y_{2}
\end{array} \right) =
\left( \begin{array}{c}
      x_{1} y_{1} \\
      x_{1} y_{2} \\
	  \hline 
	  x_{2} y_{1} \\ 
	  x_{2} y_{2}
\end{array} \right) \,.
\label{tensorvectors}
\ee 

This is a good place to introduce the very important permutation matrix
${\cal P}$, which is defined by
\be
{\cal P} = \left( \begin{array}{cc|cc}
 1  &  &  & \\ 
 &  &  1  & \\
\hline
 &  1  &  &  \\
 &  &  &  1 
\end{array} \right) \,.
\ee 
In view of Eq. (\ref{tensorvectors}),
\be
{\cal P} \left( x \otimes y \right) = \left( \begin{array}{c}
      x_{1} y_{1} \\
      x_{2} y_{1} \\
	  \hline 
	  x_{1} y_{2} \\ 
	  x_{2} y_{2}
\end{array} \right) = \left( \begin{array}{c}
      y_{1} \\
      y_{2}
\end{array} \right) \otimes
\left( \begin{array}{c}
      x_{1} \\
      x_{2}
\end{array} \right) = y \otimes x 
\,,
\ee 
and so indeed ${\cal P}$ permutes the factors in the tensor product. 
Evidently, ${\cal P}^{2} = \id \otimes \id$. Note also that
\be
{\cal P}\ \vec\sigma_{1}\ {\cal P} = \vec\sigma_{2} \,, \qquad 
{\cal P}\ \vec\sigma_{2}\ {\cal P} = \vec\sigma_{1} \,.
\ee 

The two-site Heisenberg spin Hamiltonian is \footnote{We subtract a 
constant (proportional to the identity matrix) for later convenience.}
\be
H_{12} &=& {J\over 4} \left( \vec \sigma_{1} \cdot  \vec \sigma_{2} 
- \id \otimes \id \right)
= {J\over 4} \left(
\sigma_{1}^{x} \sigma_{2}^{x} + \sigma_{1}^{y} \sigma_{2}^{y} + 
\sigma_{1}^{z} \sigma_{2}^{z} - \id \otimes \id \right) \non \\
&=& {J\over 4} \left(
\sigma^{x} \otimes \sigma^{x} +  \sigma^{y} \otimes \sigma^{y}+
 \sigma^{z} \otimes \sigma^{z} - \id \otimes \id \right) \non \\ 
&=& {J\over 2} \left( \begin{array}{cc|cc}
 0 &  0 & 0  & 0 \\ 
 0 & -1 & 1  & 0 \\
\hline
 0 &  1 & -1 & 0 \\
 0 &  0 &  0 & 0 
\end{array} \right) \,.
\label{two-site}
\ee 
We observe that
\be
H_{12} =  {J\over 2} \left( {\cal P} - \id \otimes \id \right) \,.
\label{two-site-perm}
\ee
That is, apart from multiplicative and additive constants, the 
two-site Hamiltonian is given by the permutation matrix.

We consider now the problem of diagonalizing the Hamiltonian,
\be
H_{12} | \psi \rangle = E | \psi \rangle \,.
\ee
One can verify by inspection that the solution is given by
\be
| \psi_{(1\,, 1)} \rangle &=& \left( \begin{array}{c}
      1 \\
      0 \\
	  \hline 
	  0 \\ 
	  0
\end{array} \right) = \left( \begin{array}{c}
      1 \\
      0
\end{array} \right) \otimes
\left( \begin{array}{c}
      1 \\
      0
\end{array} \right) \,, \non \\ 
| \psi_{(1\,, 0)} \rangle &=& \left( \begin{array}{c}
      0 \\
      1 \\
	  \hline 
	  1 \\ 
	  0
\end{array} \right) = \left( \begin{array}{c}
      1 \\
      0
\end{array} \right) \otimes
\left( \begin{array}{c}
      0 \\
      1
\end{array} \right) +
 \left( \begin{array}{c}
      0 \\
      1
\end{array} \right) \otimes
\left( \begin{array}{c}
      1 \\
      0
\end{array} \right) 
\,, \non \\ 
| \psi_{(1\,, -1)} \rangle &=& \left( \begin{array}{c}
      0 \\
      0 \\
	  \hline 
	  0 \\ 
	  1
\end{array} \right) = \left( \begin{array}{c}
      0 \\
      1
\end{array} \right) \otimes
\left( \begin{array}{c}
      0 \\
      1
\end{array} \right) \,, \non \\
| \psi_{(0\,, 0)} \rangle &=& \left( \begin{array}{c}
      0 \\
      1 \\
	  \hline 
	  -1 \\ 
	  0
\end{array} \right) = \left( \begin{array}{c}
      1 \\
      0
\end{array} \right) \otimes
\left( \begin{array}{c}
      0 \\
      1
\end{array} \right) -
 \left( \begin{array}{c}
      0 \\
      1
\end{array} \right) \otimes
\left( \begin{array}{c}
      1 \\
      0
\end{array} \right) 
\,. 
\ee 
The first three states have the same energy $E=0$, and the last state has 
energy $E=-J$.

This result can be easily understood from the $su(2)$ symmetry of the 
model.  Indeed, the total spin operators
\be
\vec S \equiv {1\over 2}\left( \vec\sigma \otimes \id + \id \otimes 
\vec\sigma \right)
\ee
are generators of a {\it reducible} 4-dimensional representation of 
$su(2)$.  That is, there exists a unitary matrix $U$ such that
\be
U\ \vec S\ U^{\dagger} = \left( \begin{array}{c|c}
 \vec S_{(S=1)} & \\ 
 \hline
 & \vec S_{(S=0)} 
\end{array} \right) \,,
\ee 
where $\vec S_{(S=1)}$ and $\vec S_{(S=0)}$ generate irreducible 
representations of dimension 3 and 1, respectively. Moreover,
\be
\vec S^{2} &=& {1\over 4}\left(  \vec\sigma \otimes \id + \id \otimes 
\vec\sigma \right)^{2} \non \\
&=& {1\over 4}\left(  \vec\sigma^{2} \otimes \id + 2 \vec\sigma 
\otimes \vec\sigma + \id \otimes \vec\sigma^{2} \right) \non \\
&=& {1\over 2} \vec\sigma \otimes \vec\sigma 
+ {3\over 2} \id \otimes \id \,,
\ee
since $\vec\sigma^{2} = 3 \id$. Hence the two-site Hamiltonian can be 
expressed in terms of $\vec S^{2}$,
\be
H_{12} = {J\over 4} \left( \vec\sigma \otimes \vec\sigma  
- \id \otimes \id \right)
= {J\over 2} \left(  \vec S^{2} - 2 \id \otimes \id \right)
\,.
\ee
Recalling the well-known fact 
$\vec S^{2} | S \,, S^{z} \rangle = S ( S + 1) | S \,, S^{z} \rangle $, we 
see that
\be
H_{12} | S \,, S^{z} \rangle = {J\over 2} \left(  S (S + 1) - 2 \right) 
| S \,, S^{z} \rangle
\,, \quad  S^{z} = -S \,, \ldots \,, S \,; \quad S = 0 \,, 1 \,.
\ee
In particular, for $S=1$ the energy is $E=0$, and for $S=0$ the 
energy is $E=-J$, which is the result we found earlier.

Note that for $J > 0$ (antiferromagnetic), the ground state is a spin 
singlet ($S=0$) state; while for $J < 0$ (ferromagnetic), there is a 
degenerate ground state.

\subsection{$N$ sites}

For the $N$-site problem, the basic observables are $\vec \sigma_{n} \,, 
\quad n = 1 \,, 2 \,, \ldots \,, N$, defined by
\be
\vec \sigma_{n} = 
\stackrel{\stackrel{1}{\downarrow}}{\id} 
\otimes \cdots \otimes \id \otimes 
\stackrel{\stackrel{n}{\downarrow}}{\vec \sigma} 
\otimes \id \otimes \cdots \otimes 
\stackrel{\stackrel{N}{\downarrow}}{\id} \,.
\ee
These are operators on
\be
\stackrel{\stackrel{1}{\downarrow}}{V} \otimes \cdots \otimes 
\stackrel{\stackrel{n}{\downarrow}}{V} \otimes \cdots \otimes 
\stackrel{\stackrel{N}{\downarrow}}{V} 
\label{VN/chain}
\ee
which act nontrivially on the $n^{th}$ space, and trivially on the 
rest.  

There are two possible topologies for a one-dimensional chain: open or 
closed.  Correspondingly, there are two Heisenberg spin Hamiltonians
describing nearest-neighbor interactions:
\be
H =  \sum_{n=1}^{N-1} H_{n \,, n+1}
\qquad \mbox{(open)}
\ee 
or 
\be
H =  \sum_{n=1}^{N-1} H_{n \,, n+1} + H_{N \,, 1} 
\qquad \mbox{(closed)} 
\label{Heisenberg}
\ee 
where the two-site Hamiltonian is
\be
H_{ij} = {J\over 4} \left( 
\vec \sigma_{i} \cdot \vec \sigma_{j} - \id^{\otimes N} \right) 
\label{two-site-general}
\,.
\ee 
For simplicity, we shall henceforth focus on the closed chain.

How can one diagonalize the Hamiltonian? Using a computer is not 
practical for large values of $N$. Indeed, $H$ is a $2^{N} \times 
2^{N}$ matrix. Moreover, direct numerical diagonalization will give 
all eigenstates, while often one is interested in only low-lying states.

A very elegant alternative approach which overcomes (at least in part) 
both of these difficulties is the Bethe Ansatz, of which there are 
several variants.  We focus here on the algebraic Bethe Ansatz.  
\footnote{The method pioneered by Bethe \cite{bethe} is now known as 
coordinate Bethe Ansatz.  The algebraic Bethe Ansatz was developed in 
St.  Petersburg \cite{faddeev/takhtajan1} - \cite{kbi}.  It is perhaps 
not as powerful as coordinate Bethe Ansatz, but it is more 
transparent.  Other related approaches include analytic Bethe Ansatz 
\cite{reshetikhin} and Baxter's $Q$-operator method \cite{baxter}.  A 
rather different approach, which is formulated directly in the 
thermodynamic limit and avoids Bethe Ansatz, is being developed in 
Kyoto \cite{jimbo/miwa}.} An essential element of this approach is 
a matrix $R(\lambda)$  that is a solution of the Yang-Baxter equation, 
which we discuss in Section 3.  As we shall see in Section 4, the two-site 
Hamiltonian (\ref{two-site-general}) can be expressed in terms 
of the $R$ matrix. The fact that the $R$ matrix satisfies the 
Yang-Baxter equation leads to the ``integrability''of the model, i.e., 
that the model can be solved by Bethe Ansatz.

\section{Yang-Baxter equation}

We consider the $R$ matrix \footnote{The notation 
\cite{kulish/sklyanin2} which we are using, while widely used, has not 
been universally adopted.  In particular, some authors (e.g., 
\cite{faddeev/takhtajan1}, \cite{faddeev/takhtajan2}, \cite{kbi}) call 
$R(\lambda)$ what we call $\check R(\lambda) \equiv {\cal P} 
R(\lambda)$.  Given an $R$ matrix, one can easily determine which of 
the two conventions is being used by evaluating it at $\lambda=0$: 
for us $R(0) \sim {\cal P}$, while for them $R(0) \sim \id \otimes 
\id$.  Of course, their Yang-Baxter equation looks somewhat different 
from ours.}
\be
R(\lambda) &=& \lambda \id \otimes \id  + i {\cal P} \non \\ 
&=& \left( \begin{array}{cc|cc}
 \lambda + i   &  &   & \\ 
 &  \lambda   & i & \\
\hline
 &  i  &  \lambda  &  \\
 &  &  &   \lambda + i 
\end{array} \right) =
\left( \begin{array}{cc|cc}
 a  &  &   & \\ 
 &  b  & c & \\
\hline
 &  c  &  b &  \\
 &  &  &  a
\end{array} \right) \,,
\label{Rmatrix}
\ee
where
\be
a =  \lambda + i  \,, \qquad b=  \lambda \,, \qquad c = i \,. 
\label{abc}
\ee
We regard $R(\lambda)$ as an operator which acts on $V \otimes V$. 
The variable $\lambda$ is called the spectral parameter.

We wish to show that this matrix is a solution to the so-called 
Yang-Baxter equation.  \footnote{For the case of the simple $R$ matrix 
(\ref{Rmatrix}), the approach we shall follow is unnecessarily 
tedious -- there are much faster ways to show that this matrix is a 
solution of the Yang-Baxter Eq. (\ref{YB}).  However, our approach has the 
benefit of being explicit and straightforward, and it works for any 
case.} To this end, we define $R_{12}(\lambda)$ by
\be
R_{12}(\lambda) &=& R(\lambda) \otimes \id 
= \left( \begin{array}{cc|cc}
 a  &  &   & \\ 
 &  b  & c & \\
\hline
 &  c  &  b &  \\
 &  &  &  a
\end{array} \right) \otimes 
\left( \begin{array}{cc}
	               1  & 0 \\
				   0  & 1 
\end{array} \right) \non \\
&=& \left( \begin{array}{cc|cc|cc|cc}
 a  &  &   &  &  &   &  & \\ 
 &  a  &   &  &  &   &  & \\ 
 \hline
 &  &  b   &  &  c  &   &   \\ 
 &  &  &   b  &  &  c   &   \\ 
 \hline
 &  &  c   &  &  b  &   &   \\ 
 &  &  &   c  &  &  b   &  \\
 \hline
 &  &  &   &  &  &   a  & \\ 
 &  &  &   &  &  &   &  a
\end{array} \right) \,.
\label{R12}
\ee
Evidently, this is an operator on $V\otimes V \otimes V$, which acts 
nontrivially on the first and second spaces, and trivially on the third. 

Similarly, we define $R_{23}(\lambda)$ by
\be
R_{23}(\lambda) &=& \id \otimes R(\lambda) 
= \left( \begin{array}{cc}
	               1  & 0 \\
				   0  & 1 
\end{array} \right) \otimes 
\left( \begin{array}{cc|cc}
 a  &  &   & \\ 
 &  b  & c & \\
\hline
 &  c  &  b &  \\
 &  &  &  a
\end{array} \right)  \non \\
&=& \left( \begin{array}{cc|cc|cc|cc}
 a  &  &   &   &  &   &  &    \\ 
 &  b  &   c   &  &   &  &  & \\ 
 \hline
 &  c  &   b   &  &   &  &  & \\ 
 &     &   &   a  &   &  &  & \\ 
 \hline
 &     &   &   &  a  &  &   &     \\ 
 &     &   &   &  &  b  &   c   &  \\
 \hline
 &     &   &   &  &  c  &   b   & \\ 
 &     &   &   &  &  &  &   a
\end{array} \right) \,.
\ee 
This operator acts nontrivially on the second and third spaces, and 
trivially on the first.

Finally, we wish to define $R_{13}(\lambda)$, which acts nontrivially on the 
first and third spaces, and trivially on the second. This requires only
slightly more effort:
\be
R_{13}(\lambda) = {\cal P}_{23}\ R_{12}(\lambda)\ {\cal P}_{23} \,,
\ee
where $R_{12}(\lambda)$ is given by Eq. (\ref{R12}), and ${\cal P}_{23}$
is given by
\be
{\cal P}_{23} &=& \id \otimes {\cal P}
= \left( \begin{array}{cc}
	               1  & 0 \\
				   0  & 1 
\end{array} \right) \otimes \left( \begin{array}{cc|cc}
 1  &  &  & \\ 
 &  &  1  & \\
\hline
 &  1  &  &  \\
 &  &  &  1 
\end{array} \right) \non \\
&=& \left( \begin{array}{cc|cc|cc|cc}
 1  &  &   &   &  &   &  &    \\ 
 &  &  1   &   &  &   &  &    \\ 
 \hline
 &  1  &   &   &  &   &  &   \\ 
 &     &   &   1  &   &  &  & \\ 
 \hline
 &     &   &   &  1  &  &   &     \\ 
 &     &   &   &  &  &  1   &    \\
 \hline
 &     &   &   &  &  1  &   &    \\ 
 &     &   &   &  &  &  &   1
\end{array} \right) \,.
\ee 
One can now easily obtain
\be
R_{13}(\lambda) = \left( \begin{array}{cc|cc|cc|cc}
 a   &   &      &   &   &   &   &    \\ 
 &   b   &   &   &  c  &   &   &    \\ 
 \hline
 &   &   a   &  &   &   &  &   \\ 
 &   &   &   b  &   &   & c    & \\ 
 \hline
 &   c   &   &   &  b  &   &  &        \\ 
 &   &   &   &   &  a  &     &    \\
 \hline
 &   &   &   c  &  &  & b  &    \\ 
 &   &   &   &  &  &  &   a
\end{array} \right) \,.
\ee 

The Yang-Baxter equation is
\be
R_{12}(\lambda-\lambda')\ R_{13}(\lambda)\ R_{23}(\lambda')
= R_{23}(\lambda')\ R_{13}(\lambda)\ R_{12}(\lambda-\lambda') \,.
\label{YB}
\ee
It is now a tedious but straightforward exercise in matrix 
multiplication and algebra to show that this relation is satisfied.  
With the help of a symbolic manipulation program such as Mathematica, 
the task can be accomplished in minutes. \footnote{The $R$ matrix 
(\ref{Rmatrix}) is the simplest solution of the Yang-Baxter equation. 
Many more solutions are known. See, e.g., \cite{kulish/sklyanin2}, 
\cite{jimbo}.}

\section{Transfer matrix}

In this Section, we show that the two-site Heisenberg Hamiltonian can 
be expressed in terms of the $R$ matrix, and that the model is 
integrable by virtue of the fact that this matrix is a solution of the 
Yang-Baxter equation.  However, it is more convenient to work 
``backwards'': using the $R$ matrix, we first construct the so-called 
transfer matrix -- a one-parameter commutative family of operators 
acting on the space of states of the Heisenberg spin chain.  We then 
verify that the Heisenberg Hamiltonian is among this family of 
commuting operators.
 
The key step in this program is to introduce the so-called $L$ operators 
\footnote{The shift in $\lambda $ is made in order that the Bethe 
Ansatz equations (see Eq.  (\ref{BAE}) below) have a symmetric form.  
This shift is not essential.}
\be
L_{0 n}(\lambda) &=& R_{0 n}(\lambda - {i\over 2})  \non \\
&=& \left( \begin{array}{cc}
\alpha_{n}         & \beta_{n}  \\
\gamma_{n}         & \delta_{n}
\end{array} \right) 
\,, \qquad n = 1 \,, 2 \,, \ldots \,, N \,,
\ee  
which act on so-called auxiliary ($0$) and quantum ($n$) spaces.  
That is, $L_{0 n}$ is an operator on
\be
\stackrel{\stackrel{0}{\downarrow}}{V} \otimes 
\stackrel{\stackrel{1}{\downarrow}}{V} \otimes \cdots \otimes 
\stackrel{\stackrel{n}{\downarrow}}{V} \otimes \cdots \otimes 
\stackrel{\stackrel{N}{\downarrow}}{V} 
\ee
which acts nontrivially on the $0^{th}$ and $n^{th}$ spaces, and 
trivially on the rest. Moreover,
\be
\alpha_{n} &=& 
\stackrel{\stackrel{1}{\downarrow}}{\id} 
\otimes \cdots \otimes \id \otimes 
\stackrel{\stackrel{n}{\downarrow}}{\alpha} 
\otimes \id \otimes \cdots \otimes 
\stackrel{\stackrel{N}{\downarrow}}{\id} \,, \non  \\
\beta_{n} &=& \id \otimes \cdots \otimes \id \otimes 
{\beta} \otimes \id \otimes \cdots \otimes \id \,, \non \\
\gamma_{n} &=& \id \otimes \cdots \otimes \id \otimes 
{\gamma} \otimes \id \otimes \cdots \otimes \id \,, \non \\
\delta_{n} &=& \id \otimes \cdots \otimes \id \otimes 
{\delta} \otimes \id \otimes \cdots \otimes \id \,, 
\ee
where
\be
\alpha &=& \left(\begin{array}{cc}
                \lambda+{i\over 2} & 0 \\
				0 & \lambda-{i\over 2}
			\end{array} \right) \,, \qquad 
			\beta = \left(\begin{array}{cc}
                0 & 0 \\
				i & 0
			\end{array} \right) \,, \non \\
\gamma &=& \left(\begin{array}{cc}
                0 & i \\
                0 & 0
			\end{array} \right) \,, \qquad 
			\delta = \left(\begin{array}{cc}
               \lambda-{i\over 2} & 0 \\
			   0 & \lambda+{i\over 2}
			\end{array} \right) \,.			
\ee 
Evidently, $\alpha_{n}$, $\beta_{n}$, $\gamma_{n}$, $\delta_{n}$ are 
operators on
\be
\stackrel{\stackrel{1}{\downarrow}}{V} \otimes \cdots \otimes 
\stackrel{\stackrel{n}{\downarrow}}{V} \otimes \cdots \otimes 
\stackrel{\stackrel{N}{\downarrow}}{V} 
\label{VN}
\ee
which act nontrivially on the $n^{th}$ space, and trivially on the 
rest.  The vector space (\ref{VN}) is precisely the space of states of 
the Heisenberg quantum spin chain (\ref{VN/chain}).  Note that the 
operators at site $n$ commute with the operators at any other site $n' 
\ne n$.

The algebra of the operators $\alpha_{n}$, $\beta_{n}$, $\gamma_{n}$, 
$\delta_{n}$ is encoded in the relation
\be
R_{0 0'}(\lambda-\lambda')\ L_{0 n}(\lambda)\ L_{0' n}(\lambda')
= L_{0' n}(\lambda')\ L_{0 n}(\lambda)\ R_{0 0'}(\lambda-\lambda') \,,
\label{Lalgebra}
\ee 
which immediately follows from the Yang-Baxter Eq. (\ref{YB}) upon 
replacing $1 \rightarrow 0$, $2 \rightarrow 0'$, and
$3 \rightarrow n$, and also making the shifts 
$\lambda \rightarrow \lambda - {i\over 2}$ and  
$\lambda' \rightarrow \lambda' - {i\over 2}$. 

The monodromy matrix $T_{0}(\lambda)$ is defined as the 
following product of $L$ operators \footnote{As is customary, we often 
suppress the quantum-space subscripts.}
\be
T_{0}(\lambda) &=& L_{0 N}(\lambda) \cdots L_{0 1}(\lambda) \non \\
&=& \left(\begin{array}{cc}
                \alpha_{N} & \beta_{N} \\
				\gamma_{N} & \delta_{N}
			\end{array} \right) \cdots 
 \left(\begin{array}{cc}
                \alpha_{1} & \beta_{1} \\
				\gamma_{1} & \delta_{1}
			\end{array} \right)	
\label{monodromy}			
\,.
\ee
The monodromy matrix obeys the fundamental relation
\be
R_{0 0'}(\lambda-\lambda')\ T_{0}(\lambda)\ T_{0'}(\lambda')
= T_{0'}(\lambda')\ T_{0}(\lambda)\ R_{0 0'}(\lambda-\lambda') \,.
\label{fundamental}
\ee 
We prove this result for $N=2$. The left hand side (LHS) is then
\be
LHS &=& R_{0 0'}(\lambda-\lambda')\ L_{0 2}(\lambda)\ L_{0 1}(\lambda)\ 
L_{0' 2}(\lambda')\ L_{0' 1}(\lambda') \non \\
&=& R_{0 0'}(\lambda-\lambda')\ L_{0 2}(\lambda)\ L_{0' 2}(\lambda')\
L_{0 1}(\lambda)\ L_{0' 1}(\lambda') \non \\
&=& L_{0' 2}(\lambda')\ L_{0 2}(\lambda)\ R_{0 0'}(\lambda-\lambda')\ 
L_{0 1}(\lambda)\ L_{0' 1}(\lambda') \non \\
&=& L_{0' 2}(\lambda')\ L_{0 2}(\lambda)\ L_{0' 1}(\lambda')\ 
L_{0 1}(\lambda)\ R_{0 0'}(\lambda-\lambda') \non \\
&=& L_{0' 2}(\lambda')\ L_{0' 1}(\lambda')\ L_{0 2}(\lambda)\ 
L_{0 1}(\lambda)\ R_{0 0'}(\lambda-\lambda') = RHS \,. 
\ee 
In passing to the second line, we have used the fact that $L_{0 n}$ 
commutes with $L_{0' n'}$ for $n \ne n'$, and in passing to the third 
line we have used the fact (\ref{Lalgebra}). This proof immediately 
extends to the case of general $N$.

The transfer matrix $t(\lambda)$ is defined by
tracing over the auxiliary space
\be
t(\lambda) = \tr_{0} T_{0}(\lambda) \,.
\label{transfer}
\ee
It acts only on the quantum spaces (\ref{VN}). The transfer matrix
constitutes a one-parameter family of commuting operators
\be
\left[ t(\lambda)\,, t(\lambda') \right] = 0 \,.
\label{commutativity}
\ee
Indeed, multiplying the fundamental relation (\ref{fundamental}) on 
the right by the inverse $R$ matrix, and taking the trace over both 
auxiliary spaces $0$ and $0'$, we obtain
\be
\tr_{0 0'} R_{0 0'}(\lambda-\lambda')\ T_{0}(\lambda)\ T_{0'}(\lambda')\ 
R_{0 0'}(\lambda-\lambda')^{-1}
= \tr_{0 0'} T_{0'}(\lambda')\ T_{0}(\lambda)\  \,.
\ee 
Using the cyclic property of the trace, we obtain
\be
\tr_{0 0'} T_{0}(\lambda)\ T_{0'}(\lambda') 
= \tr_{0 0'} T_{0'}(\lambda')\ T_{0}(\lambda)  \,,
\ee 
which implies the result (\ref{commutativity}).

The transfer matrix contains the momentum operator $P$ of the 
(closed) quantum spin chain. Specifically,
\be
P={1\over i}\log \ i^{-N} t({i\over 2})  \,. 
\label{momentum}
\ee
We show this for the case $N=2$. We first observe
\be
t({i\over 2}) = \tr_{0} T_{0}({i\over 2}) = 
\tr_{0} R_{0 2}(0)\ R_{0 1}(0) 
 = i^{2} \tr_{0} {\cal P}_{0 2}\ {\cal P}_{0 1} = -{\cal P}_{12} 
\label{tat}
\,,
\ee
where we have used the fact $R(0)=i{\cal P}$ (see Eq.  
(\ref{Rmatrix})) and the identity 
$\tr_{0} A_{0} {\cal P}_{0 n} = A_{n}$.  It follows that
\be
t({i\over 2})\  X_{1}\ t({i\over 2})^{-1} = 
{\cal P}_{12}\ X_{1}\ {\cal P}_{12} = X_{2} \,,
\ee
where $X_{n}$ is any operator at site $n$. One can show for general $N$
in similar fashion that $t({i\over 2})$ is the one-site shift operator,
\be
t({i\over 2})\  X_{n}\ t({i\over 2})^{-1} = X_{n+1} \,, \qquad 
n = 1 \,, \ldots \,, N \,.
\ee
We recall from elementary quantum mechanics that
\be
e^{i a P}\ X\ e^{-i a P} = X + a \,,
\ee
where $P$ and $X$ are the momentum and position operators, and $a$ is 
a c-number. By analogy, we define the momentum operator for a spin 
chain by $e^{iP} = i^{-N} t({i\over 2})$, which implies (\ref{momentum}).

The transfer matrix also contains a closed-chain Hamiltonian 
\be
H = {J\over 2}\left( i {d\over d \lambda} \log 
t(\lambda)\Big\vert_{\lambda = {i\over 2}} 
- N \id^{\otimes N} \right) 
= \sum_{n=1}^{N-1} H_{n\,, n+1} + H_{N \,,1} \,,
\label{Hamiltonian}
\ee 
with the two-site Hamiltonian
\be
H_{ij} = {J\over 2} \left( {\cal P}_{ij}\ R_{ij}'(0) - 
\id^{\otimes N} \right) \,,
\label{two-site-R}
\ee 
where the prime denotes differentiation with respect to $\lambda$.
This is easily verified for $N=2$ using (\ref{tat}) and also
\be
t'({i\over 2}) = \tr_{0} R'_{0 2}(0)\ R_{0 1}(0) + 
\tr_{0} R_{0 2}(0)\ R'_{0 1}(0) = 2 i R'_{1 2}(0) \,. 
\ee
Moreover, keeping in mind the result (\ref{two-site-perm}) and the
expression (\ref{Rmatrix}) for our $R$ matrix, we see that the two-site 
Hamiltonian (\ref{two-site-R}) coincides with that of the Heisenberg 
chain (\ref{two-site-general}). In short, the Hamiltonian 
(\ref{Hamiltonian}) is precisely the Heisenberg Hamiltonian 
(\ref{Heisenberg}).

Combining the results (\ref{commutativity}) and (\ref{Hamiltonian}), 
we see that the Hamiltonian commutes with the transfer matrix,
\be
\left[ H\,, t(\lambda) \right] = 0 \,.
\ee
This relation together with (\ref{commutativity}) signal the 
``integrability'' of the model.

\section{Algebraic Bethe Ansatz}

We finally turn to the problem of diagonalizing the Hamiltonian. In 
fact, we shall diagonalize the transfer matrix. The strategy is to 
identify certain creation operators with which to construct states, 
as one does for the simple harmonic oscillator. These operators, however, 
depend on parameters; the states are eigenstates of the transfer 
matrix if the parameters are solutions of a set of equations, namely 
the Bethe Ansatz equations.

The monodromy matrix $T_{0}(\lambda)$ is a $2 \times 2$ matrix in the 
auxiliary space, whose elements are operators on the quantum space 
$V^{\otimes N}$ which we denote as follows
\be
T_{0}(\lambda) =
\left( \begin{array}{cc}
A(\lambda) & B(\lambda)  \\
C(\lambda) & D(\lambda)
\end{array} \right) 
\,.
\ee
A set of algebraic relations among these four operators is encoded in 
the fundamental relation (\ref{fundamental}).  The relations include
\be
\left[ B(\lambda) \,, B(\lambda') \right] = 0 \,, \non 
\ee
\be 
A(\lambda)\ B(\lambda') &=& {a(\lambda' - \lambda)\over b(\lambda' - \lambda)}
B(\lambda')\ A(\lambda) - 
{c(\lambda' - \lambda)\over b(\lambda' - \lambda)} B(\lambda)\ A(\lambda')
\,, \non \\
D(\lambda)\ B(\lambda') &=& {a(\lambda - \lambda')\over b(\lambda - \lambda')}
B(\lambda')\ D(\lambda) - 
{c(\lambda - \lambda')\over b(\lambda - \lambda')} B(\lambda)\ 
D(\lambda') \,,
\label{algebra}
\ee
where $a \,, b \,, c$ are given by Eq. (\ref{abc}).

Let $\omega_{+}$ be the ferromagnetic vacuum state with all spins up,
\be
\omega_{+} =  \underbrace{{1 \choose 0}
\otimes \cdots \otimes {1 \choose 0}}_{N} \,,
\ee
which is an eigenstate of $A(\lambda)$ and $D(\lambda)$,
and which is annihilated by $C(\lambda)$,
\be
A(\lambda)\ \omega_{+} = \left( \lambda + {i\over 2} \right)^{N} 
\omega_{+} \,, \qquad 
D(\lambda)\ \omega_{+} = \left( \lambda - {i\over 2} \right)^{N} 
\omega_{+} \,, \qquad 
C(\lambda)\ \omega_{+} = 0 \,. 
\label{properties}
\ee 
We use the operators $B(\lambda)$ as creation operators to construct 
the so-called Bethe state
\be
|\lambda_{1} \,, \ldots \,, \lambda_{M} \rangle =
B(\lambda_{1}) \cdots B(\lambda_{M})\ \omega_{+} \,.
\label{vec}
\ee
Using the algebraic relations (\ref{algebra}) and the properties 
(\ref{properties}) of the ferromagnetic vacuum state, it can be shown 
that the Bethe state is an eigenstate of the transfer matrix 
$t(\lambda) = A(\lambda) + D(\lambda)$,
\be
t(\lambda)\ |\lambda_{1} \,, \ldots \,, \lambda_{M} \rangle =
\Lambda(\lambda \,; \lambda_{1} \,, \ldots \,, \lambda_{M})\  
|\lambda_{1} \,, \ldots \,, \lambda_{M} \rangle
\ee 
with the eigenvalue
\be
\Lambda(\lambda \,; \lambda_{1} \,, \ldots \,, \lambda_{M}) 
= \left( \lambda + {i\over 2} \right)^{N}
\prod_{\alpha=1}^{M} 
{ \lambda - \lambda_{\alpha} - i 
\over 
  \lambda - \lambda_{\alpha}  } 
+ \left( \lambda - {i\over 2}  \right)^{N}
\prod_{\alpha=1}^{M} 
{ \lambda - \lambda_{\alpha} + i 
\over 
  \lambda - \lambda_{\alpha}  } 
\,,
\label{transfereigen}
\ee
if $\{ \lambda_{1} \,, \ldots \,, \lambda_{M} \}$ are distinct and obey 
the Bethe Ansatz equations
\be
\left(  {\lambda_{\alpha} + {i\over 2} 
\over   \lambda_{\alpha} - {i\over 2}} \right)^{N} 
= \prod_{\scriptstyle{\beta=1}\atop \scriptstyle{\beta \ne \alpha}}^M 
{\lambda_{\alpha} - \lambda_{\beta} + i 
\over 
 \lambda_{\alpha} - \lambda_{\beta} - i }
\,, \qquad \alpha = 1 \,, \cdots \,, M \,. 
\label{BAE}
\ee
The proof is immediate for $N=1$, and is not much more difficult for 
general $N$.  It is well-described in many papers (see, e.g., 
\cite{faddeev/takhtajan1},\cite{kulish/sklyanin1},\cite{kbi}), so 
we shall not repeat it here.

In particular, one can now obtain the momentum and energy 
eigenvalues using Eqs.  (\ref{momentum}),
(\ref{Hamiltonian}), and (\ref{transfereigen}), 
\be
P = {1\over i} \sum_{\alpha=1}^{M} 
\log \left( {\lambda_{\alpha} + {i\over 2} 
\over \lambda_{\alpha} - {i\over 2}} \right)
\quad (\mbox{mod } 2 \pi)
\,, \qquad 
E = - {J\over 2} \sum_{\alpha=1}^{M} {1\over \lambda_{\alpha}^{2} + {1\over 4}}
\,.
\label{energy/momentum}
\ee 

An important feature of the Heisenberg spin chain is its $su(2)$ 
symmetry.  One can show \cite{faddeev/takhtajan2} that the transfer 
matrix commutes with the total spin operators,
\be
\left[ \vec S \,, t(\lambda) \right] = 0 \,, \qquad 
\vec S = {1\over 2} \sum_{n=1}^{N} \vec \sigma_n \,.
\label{spin}
\ee 
Moreover, the Bethe states (\ref{vec}) are $su(2)$ highest weight states
\be
S^{+}\ |\lambda_{1} \,, \ldots \,, \lambda_{M} \rangle = 0 \,, \qquad
S^{\pm} = S^{x} \pm i S^{y} \,, 
\ee
and are eigenstates of $S^{z}$ with eigenvalue 
\be 
S^{z}={N\over 2} - M \,.
\label{spineigenvalue}
\ee 
Since $S = S^{z} \ge 0$, it follows that $M \le N/2$. Lower weight 
states are obtained by acting on the Bethe states with the lowering 
operator $S^{-}$.

\section{Now the hard work begins}

The $S = 1/2$ isotropic Heisenberg spin chain which we have discussed 
is the simplest integrable magnetic chain.  Many generalizations are 
possible.  For instance, using appropriate $R$ matrices, one can 
construct anisotropic (XXZ and XYZ) chains, chains with spins in 
higher-dimensional representations ($S= 1 \,, 3/2 \,, \ldots$) or with 
combinations of different spins, and chains with spins in 
representations of higher-rank algebras ($su({\cal N}) \,, \ldots $).  
Using also so-called $K$ matrices, one can construct integrable open 
spin chains \cite{sklyanin}.  Bethe Ansatz equations have been 
obtained for many such models.  \footnote{Numerous additional 
references can be found in \cite{kbi}.}

Obtaining Bethe Ansatz equations (BAE) for an integrable model is in a 
sense ``half'' the solution of the model -- the easier half at that!  
The difficulty is that quantities of physical interest (e.g., energy) 
are expressed in terms of solutions of BAE; and unfortunately, BAE are 
generally hard to solve.  Various clever methods have been developed 
for extracting from these equations information about the 
thermodynamic ($N \rightarrow \infty$) limit.  Invariably the methods 
involve some assumptions about the nature of the solutions, such as 
the string hypothesis (see, e.g., \cite{faddeev/takhtajan2}).  The 
types of results which have been obtained include spectrum of 
low-lying states, scattering matrices, finite-size corrections, and 
thermodynamics (finite temperature and magnetic field).  Progress has 
been made on the computation of correlation functions.  See, e.g., 
\cite{kbi} and other contributions to these Proceedings.

\section*{Acknowledgments}

I thank L.  Castellani for his kind invitation to this stimulating 
conference, and for encouraging me to prepare this review.  I am also 
grateful to O.  Alvarez and F.  Essler for their comments on the 
manuscript.  This work was supported in part by the National Science 
Foundation under Grant PHY-9870101.

\end{document}